\documentclass[
  reprint,
  superscriptaddress,
  amsmath,
  amssymb,
  aps,
  prl,
]{revtex4-2}

\usepackage{amsmath,amssymb,bm,mathtools,mathrsfs}
\usepackage{graphicx}
\usepackage{microtype}
\usepackage{xcolor}
\usepackage[colorlinks=true,allcolors=black]{hyperref}
\graphicspath{{figures/}}

\newcommand{\ii}{\mathrm{i}}

\begin{document}

\title{Topological Quantum Reservoir Computing}

\author{Zhi-Lei Zhang}
\affiliation{
  Graduate School of China Academy of Engineering Physics,
  Beijing 100193, China
}

\author{Lianxiang Cui}
\email{lxcui@g.ecc.u-tokyo.ac.jp}
\affiliation{International Research Center for Neurointelligence, University of Tokyo Institutes for Advanced Study, University of Tokyo, Bunkyo-ku, Tokyo 113-0033, Japan}

\author{C. P. Sun}
\affiliation{
  Graduate School of China Academy of Engineering Physics,
  Beijing 100193, China
}

\author{Kohei Nakajima}
\affiliation{Graduate School of Information Science and Technology, University of Tokyo, Bunkyo-ku, Tokyo 113-0033, Japan}
\affiliation{International Research Center for Neurointelligence, University of Tokyo Institutes for Advanced Study, University of Tokyo, Bunkyo-ku, Tokyo 113-0033, Japan}

\author{Kazuyuki Aihara}
\affiliation{International Research Center for Neurointelligence, University of Tokyo Institutes for Advanced Study, University of Tokyo, Bunkyo-ku, Tokyo 113-0033, Japan}

\begin{abstract}
\begingroup\normalcolor
Scaling universal quantum computation and implementing quantum machine learning with protected logical qubits remain challenging. 
To harness the computational potential of existing fault-tolerant quantum platforms, we introduce topological quantum reservoir computing, a novel approach using protected degrees of freedom for storage and the host system’s intrinsic dynamics for information processing.
Using a one-dimensional Kitaev chain, we show how long-lived Majorana edge modes and fast bulk responses support these complementary functions within the same many-body system. 
We identify the tradeoffs governing access to protected memory and show how matched interfaces enable programmable writing and interactions enable autonomous information consolidation from the bulk to the edge.
These mechanisms improve performance on tasks requiring long-term retention of temporal information. 
Our results establish topology as a new resource for quantum reservoir computing and open a route toward near-term quantum computing and quantum machine learning.
\par\endgroup
\end{abstract}

\maketitle

\begingroup\normalcolor
\textit{Introduction.---}
% Quantum reservoir computing (QRC) provides a dynamical approach to quantum machine learning, using the intrinsic dynamics of a quantum many-body system to transform temporal inputs into dynamical features while training only simple linear readouts for specific tasks~\cite{jaeger2001,nakajima2020prc,fujii2017,nakajima2019}. 
% The many-body evolution acts as a physical feature map, generating a rich set of dynamical responses without training the internal dynamics of the reservoir.
% QRC has been explored across a variety of quantum platforms, including spin systems~\cite{palacios2024,hou2026}, photonic and bosonic systems~\cite{nokkala2021,dudas2023,paparelle2026}, superconducting circuits~\cite{senanian2024,suzuki2022,carles2026,kubota2023noise}, and neutral-atom arrays~\cite{araiza2022,liu2026}.
% Many temporal tasks require a reservoir to respond effectively to new inputs while retaining information about earlier inputs over much longer timescales. 
% In conventional QRC, both capabilities are governed by the same underlying dynamics, making it nontrivial to extend memory without affecting the dynamics responsible for information processing~\cite{dambre2012,martinez2021,kobayashi2026,tran2021tomography}.
Quantum error correction and topological encoding provide routes to protecting quantum information by distributing it across many physical degrees of freedom~\cite{shor1995qec,terhal2015qec,kitaev2003anyons,nayak2008tqc}. However, scaling universal quantum computation to multiple protected logical qubits remains challenging in both physical resources and control~\cite{google2025belowthreshold,butt2026measurementfree,aasen2025blueprint}. To better exploit existing technologies for quantum computing and quantum machine learning, we therefore ask: can the physical system supporting protected information itself perform computation through its intrinsic dynamics, while its protected degrees of freedom retain information generated during processing?

% Quantum systems hosting topologically protected boundary states are natural candidates for addressing this problem~\cite{su1979ssh,kitaev2001,ryu2002edge,chiu2016}. Such boundary states have been widely studied in systems including quantum Hall phases, topological insulators, and topological superconductors~\cite{fukane2008,lutchyn2010,oreg2010,leijnse2012poor,qiao2022magnetic,yue2023refined,qiao2024dressed,zhang2026dressed}. A minimal model is the one-dimensional Kitaev chain, which describes spinless fermions with \(p\)-wave pairing and supports Majorana modes localized at its two ends in the topological phase~\cite{kitaev2001,alicea2012}. The coexistence of localized edge modes and extended bulk modes offers a natural route to separating dynamical timescales within the same physical system. When used as a reservoir, the bulk can retain rich and relatively fast dynamics, while the protected edge sector can provide a much longer-lived memory coordinate. Topology therefore provides a physically distinct long-memory sector whose timescale is set independently of the bulk feature dynamics.
We explore this possibility by introducing topological quantum reservoir computing (TQRC). Quantum reservoir computing (QRC) provides a dynamical approach to quantum machine learning, using the intrinsic evolution of a quantum many-body system as a physical feature map for temporal inputs while training only a linear readout for a given task~\cite{jaeger2001,nakajima2020prc,fujii2017,nakajima2019}. QRC has been explored across a variety of platforms, including spin systems~\cite{palacios2024,hou2026}, photonic and bosonic systems~\cite{nokkala2021,dudas2023,paparelle2026}, superconducting circuits~\cite{senanian2024,suzuki2022,carles2026,kubota2023noise}, and neutral-atom arrays~\cite{araiza2022,liu2026}. Topological quantum systems offer a distinctive setting for this approach through the coexistence of extended bulk modes and protected boundary modes~\cite{su1979ssh,kitaev2001,ryu2002edge,chiu2016}. Such boundary states have been widely studied in quantum Hall systems, topological insulators, and topological superconductors~\cite{qizhang2011,fukane2008,lutchyn2010,oreg2010,qiao2022magnetic,qiao2024dressed,zhang2026dressed}. This bulk–boundary structure provides a natural route to separating dynamical timescales: bulk modes can support fast information processing, while protected boundary modes provide long-lived memory.
\par\endgroup
\begin{figure}[b]
\centering
\includegraphics[width=\columnwidth]{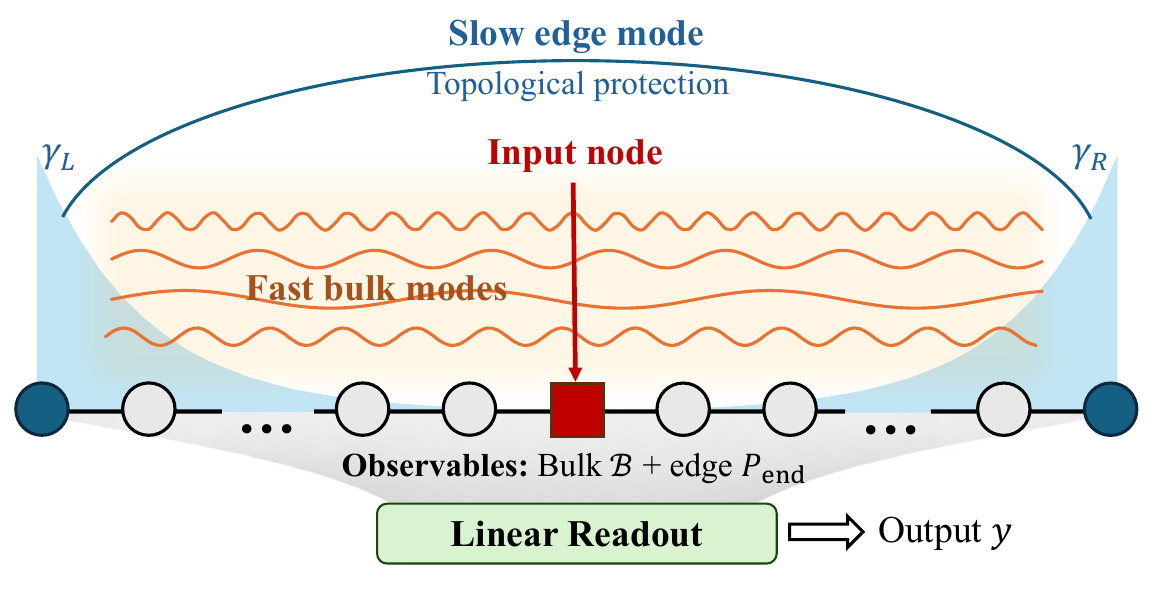}
% \caption{\textbf{Topological quantum reservoir computing architecture.}
% Temporal inputs enter through the central site of a Kitaev chain.
% Fast bulk modes generate dynamical features for short-time processing,
% while the boundary Majoranas $\gamma_L$ and $\gamma_R$ support a
% topologically protected slow edge mode for long-time information retention.
% The local bulk observables $\mathcal{B}$ and edge observable $P_{\rm end}$
% feed a linear readout that produces the output $y$.
\caption{\textbf{Topological quantum reservoir computing.}
Schematic of a Kitaev-chain reservoir.
At each cycle, the input \(u_k\) is encoded by resetting the central-site occupation. 
Fast bulk modes generate dynamical features for short-time processing, while the Majorana edge modes \(\gamma_L\) and \(\gamma_R\) support a long-lived edge parity for information retention. 
Expectation values of local Majorana bilinears \(\mathcal B\) and the nonlocal endpoint parity \(P_{\rm end}\) are supplied to a trained linear readout to produce the output \(y\).}
\label{fig:architecture}
\end{figure}

\begin{figure*}[t]
\centering
\includegraphics[width=\textwidth]{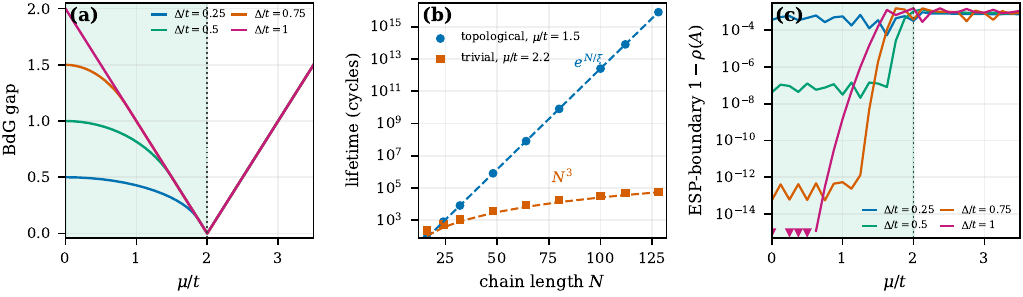}
% \caption{\textbf{Topological phase and protected edge retention.}
% (a) Thermodynamic bulk Bogoliubov--de Gennes gap of the noninteracting Kitaev chain for
% several pairing strengths $\Delta/t$.
% The gap closes at $|\mu|=2|t|$, separating the topological and trivial
% phases.
% (b) System-size dependence of the longest retention time under the
% central-reset protocol.
% In the topological phase ($\mu/t=1.5$) it grows exponentially,
% $\tau_{\rm e}\sim e^{N/\xi}$, whereas in the trivial phase
% ($\mu/t=2.2$) the longest-lived mode scales algebraically as $N^3$.
% (c) Distance to the ESP boundary $1-\rho(A)$ versus $\mu/t$ under the
% central-reset protocol for $N=32$, $t\tau=1.5$, and several $\Delta/t$.
% The exponentially slow branch occurs within the topological phase;
% downward triangles indicate values below numerical resolution.}
\caption{\textbf{Topological phase and protected edge retention.}
(a) Thermodynamic bulk Bogoliubov--de Gennes gap of the noninteracting
Kitaev chain for the indicated pairing strengths $\Delta/t$.
The gap closes at $|\mu|=2|t|$; shading in (a,c) marks the
topological phase.
(b) Lifetime versus chain length $N$ under central-site input reset,
with $\Delta=t$ and $t\tau=1.5$.
In the topological phase ($\mu/t=1.5$), the edge lifetime grows
exponentially as $\tau_{\rm e}\sim e^{N/\xi}$; in the trivial phase
($\mu/t=2.2$), the slowest-mode lifetime grows algebraically as $N^3$.
(c) Distance $1-\rho(A)$ to the echo-state-property (ESP) boundary
under the same reset protocol, with $N=32$ and $t\tau=1.5$.
Downward triangles mark values below numerical resolution.}
\label{fig:memory}
\end{figure*}

% In this Letter, we introduce topological quantum reservoir computing and realize this bulk–edge architecture in a one-dimensional Kitaev chain (Fig.~\ref{fig:architecture}).
% In the noninteracting limit, the input-driven dynamics forms an exact linear reservoir in covariance space, allowing us to derive its echo-state condition and memory function.
% We find that the protected edge mode supports exponentially long memory, while its local edge write–read amplitude is exponentially suppressed, yielding the protection–access tradeoff.
% Direct coupling of an ancillary fermion to the nonlocal topological edge mode removes this local-access penalty and turns the protected edge mode into a programmable long-term memory with an exact write–retention tradeoff. 
% When interactions are introduced, information can be transferred autonomously from the bulk into the edge sector, while the same interactions reduce edge retention, producing a loading–retention competition that extends temporal processing to longer timescales.
\begingroup\normalcolor
In this Letter, we achieve TQRC through a one-dimensional Kitaev chain~\cite{kitaev2001,alicea2012} (Fig.~\ref{fig:architecture}), whose Majorana edge modes support a long-lived parity memory alongside faster bulk dynamics. 
In the noninteracting limit, the input-driven dynamics forms an exact linear reservoir in covariance space, allowing us to derive its echo-state condition and memory function. 
% We uncover a protection–access tradeoff: the edge-memory lifetime grows exponentially with chain length under central-site input driving, while local write–read access is exponentially suppressed. 
We find that the protected edge mode supports exponentially long memory, while its local edge write–read amplitude is exponentially suppressed, yielding the protection–access tradeoff.
A matched ancillary interface removes this local-access penalty and turns the protected edge mode into a programmable memory with an exact write–retention tradeoff. 
When interactions are introduced, information can be transferred autonomously from the bulk into the edge sector, while the same interactions reduce edge retention, producing a loading–retention competition that extends temporal processing to longer timescales.
Together, these results establish topology as a resource for integrating dynamical information processing and protected long-term memory within a single quantum reservoir.
\par\endgroup

\textit{Kitaev reservoir.---}
Consider an open Kitaev chain~\cite{kitaev2001,alicea2012} of $N$ spinless fermions,
\begin{align}
H={}&-\mu\sum_{j=1}^{N}\left(n_j-\tfrac12\right)
-t\sum_{j=1}^{N-1}(c_j^\dagger c_{j+1}+\mathrm{H.c.})\nonumber\\
&+\Delta\sum_{j=1}^{N-1}(c_jc_{j+1}+\mathrm{H.c.})\nonumber\\
&+U\sum_{j=1}^{N-1}\left(n_j-\tfrac12\right)
\left(n_{j+1}-\tfrac12\right).
\label{eq:H}
\end{align}
Here \(c_j\) and \(c_j^\dagger\) annihilate and create a fermion at site \(j\), respectively, and \(n_j=c_j^\dagger c_j\) is the corresponding particle-number operator.
The parameters \(\mu\), \(t\), \(\Delta\), and \(U\) denote the chemical potential, hopping amplitude, \(p\)-wave pairing, and nearest-neighbor density interaction, respectively. 
We first consider $U=0$.  
In the thermodynamic limit, the Bogoliubov–de Gennes gap closes at \(|\mu|=2|t|\), separating the topological phase \(|\mu|<2|t|\) from the trivial phase, as shown in Fig.~\ref{fig:memory}(a). 
Introducing Majorana operators $a_j=c_j+c_j^\dagger$ and $b_j=-\ii(c_j-c_j^\dagger)$, the topological phase supports exponentially localized Majorana edge modes at the two ends of the chain, in addition to the extended bulk modes, as illustrated in Fig.~\ref{fig:architecture}~\cite{kitaev2001,alicea2012}.

To use the Kitaev chain as a reservoir, at each clock cycle $k$ we encode a scalar input $u_k$ in the occupation of the central site $q$, replacing it by $\rho_q(u_k)=(I_q+u_kP_q)/2$, with $P_q=2n_q-1$.
The remaining sites retain their reduced state, after which the full chain evolves for a fixed time $\tau$ under $H$.
The reservoir update is therefore
\begin{equation}
\rho_{k+1}
=
\mathcal{U}_U\!\left[\mathcal{R}_{u_k}(\rho_k)\right].
\label{eq:reservoir_update}
\end{equation}
Here $\mathcal{R}_{u}(\rho)=\rho_q(u)\otimes\mathrm{Tr}_q\rho$
denotes the input reset, and
$\mathcal{U}_U(X)=e^{-iH(U)\tau}Xe^{iH(U)\tau}$
the subsequent unitary evolution. 
The reservoir is read out through expectation values of local Majorana bilinears
$\mathcal{B}
=
\{\,\ii a_j b_j\,\}_{j=1}^{N}
\cup
\{\,\ii a_j a_{j+1},
\ii a_j b_{j+1},
\ii b_j a_{j+1},
\ii b_j b_{j+1}\,\}_{j=1}^{N-1}$.
To access the protected edge mode, we additionally consider the nonlocal endpoint parity $P_{\rm end}=\ii a_1 b_N$.
The expectation values of the chosen observables are collected into the feature vector $z_k$.
Here we can adopt ensemble-averaged readout over independent realizations of the same input history, each measured only at the chosen readout time, as in standard restarting protocols~\cite{fujii2017,suzuki2022,mujal2023measurement}, thereby avoiding the accumulation of measurement back action during reservoir evolution.
For a target $y_k$, only a linear readout $\hat y_k=W_{\rm out}z_k$ is trained, with $W_{\rm out}=\operatorname*{arg\,min}_{W}
\sum_k\|y_k-\hat y_k\|^2$.

\begin{figure*}[t]
\centering
\includegraphics[width=\textwidth]{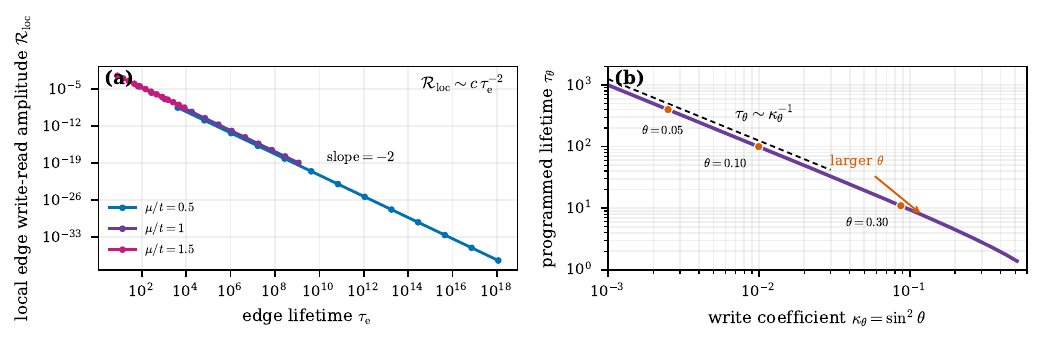}
\caption{\textbf{Protection--access tradeoff and programmable edge memory.}
(a) Local access to the protected edge mode.
For a central input and observation interface, the write--read residue
$\mathcal{R}_{\rm loc}$ decreases as the edge retention time $\tau_{\rm e}$ increases.
Eliminating the system size $N$ gives the protection--access law
$\mathcal{R}_{\rm loc}\sim\tau_{\rm e}^{-2}$ for different parameters in the
topological phase.
(b) Programmable edge memory obtained by directly coupling a fresh
ancillary fermion to the nonlocal edge fermion.
The retention time $\tau_\theta$
is shown versus the write coefficient $\kappa_\theta$.
Increasing $\theta$ enhances writing at the cost of shorter retention.
}
\label{fig:access}
\end{figure*}

When $U=0$, the Hamiltonian is quadratic and the dynamics closes exactly on the Majorana covariance matrix
$\Gamma_{\alpha\beta}=(i/2)\mathrm{Tr}\,\rho[w_\alpha,w_\beta]$,
where $w=(a_1,b_1,\ldots,a_N,b_N)$~\cite{bravyi2005}.
Under the protocol in Eq.~\eqref{eq:reservoir_update}, the covariance matrix obeys
\begin{equation}
\Gamma_{k+1}=L\Gamma_kL^{T}+u_kJ ,
\label{eq:covariance_update}
\end{equation}
where $L$ describes the one-cycle propagation of the pre-existing covariance and $J$ the covariance injected by the input.
Collecting the independent elements $\Gamma_{\alpha\beta}$ 
($\alpha<\beta$) into a reservoir state $x_k$, Eq.~\eqref{eq:covariance_update}
takes the standard linear form
\begin{equation}
x_{k+1}=Ax_k+gu_k ,
\label{eq:linear_reservoir}
\end{equation}
where \(A\) and \(g\) are the recurrence matrix and input vector, respectively.
Thus, the noninteracting Kitaev chain forms an exact linear reservoir in covariance space.
See Supplemental Material (SM)~\cite{SM}, Sec.~S1 for detailed derivations.

This exact linear form makes both stability and memory analytically accessible~\cite{haruna2019,guan2025}. 
The linear reservoir satisfies the echo-state property (ESP) if and only if $\rho(A)<1$, which guarantees that the (covariance-space) reservoir state is uniquely determined by the input history and is independent of the initial state~\cite{nokkala2021,martinez2023,kobayashi2024esp}.
For a set of observed features $z_k=Mx_k$, the response to an input applied $d$ cycles earlier is $h_d=MA^d g$. 
For independent zero-mean inputs, the optimal linear readout gives the delayed-input memory function
% \begin{equation}
% C_d
% =
% h_d^{T}
% \left(
% \sum_{j\geq0}h_jh_j^{T}
% \right)^{-1}
% h_d ,
% \label{eq:memory_function}
% \end{equation}
\(
C_d
=
h_d^{T}
\left(
\sum_{j\geq0}h_jh_j^{T}
\right)^{-1}
h_d
\), which quantifies the linearly accessible memory at delay $d$~\cite{dambre2012,cui2026rome}.

\textit{Protected edge memory and local access.—}
The topological edge mode can be extremely long lived, while its localization strongly suppresses its coupling to local bulk input and readout.
% In the topological phase, the left and right Majorana wave functions decay exponentially into the bulk as \( |\ell_j|\sim e^{-(j-1)/\xi}\) and \( |r_j|\sim e^{-(N-j)/\xi}\), where \(\xi\) is the Majorana localization length.
We denote the normalized left and right Majorana edge operators by
$\gamma_L=\sum_j \ell_j a_j$ and
$\gamma_R=\sum_j r_j b_j$.
Their wave functions decay exponentially into the bulk as
$|\ell_j|\sim e^{-(j-1)/\xi}$ and
$|r_j|\sim e^{-(N-j)/\xi}$,
where $\xi$ is the Majorana localization length.
Because the input reset acts at the center of the chain, it removes only an exponentially small fraction of the edge modes.
Accordingly, the corresponding eigenvalue \(\lambda_{\rm e}\) of the reservoir map \(A\) satisfies \(1-|\lambda_{\rm e}|\sim e^{-N/\xi}\).
The associated retention time, \(\tau_{\rm e}=-1/\ln|\lambda_{\rm e}|\), therefore grows as \(e^{N/\xi}\).

In the trivial phase, the slowest mode is an extended standing wave whose amplitude can be anomalously small at the central reset site. 
This mode has an amplitude scaling as \(N^{-3/2}\) at the reset site, so its loss per cycle scales as \(N^{-3}\).
Accordingly, the distance to the ESP boundary scales with system size as
\begin{equation}
1-\rho(A)\sim
\begin{cases}
e^{-N/\xi}, & |\mu|<2|t|,\\[2mm]
N^{-3}, & |\mu|>2|t|.
\end{cases}
\label{eq:forgetting_scaling}
\end{equation}
Figure~\ref{fig:memory}(b) confirms the exponential and algebraic size
scalings in the topological and trivial phases, respectively.
The corresponding change in $1-\rho(A)$ across the topological
transition is shown in Fig.~\ref{fig:memory}(c).

% Long retention, however, does not necessarily imply strong access to
% the stored information.
% For a local input and observation interface, the magnitude of the edge
% contribution to the impulse response can be written as
% $|h_d^{(\mathrm e)}|=R_{\rm loc}|\lambda_{\rm e}|^d$,
% where $R_{\rm loc}=|g_{\rm in}g_{\rm out}|$ and
% $g_{\rm in}$ and $g_{\rm out}$ denote the local write and read amplitudes
% of the edge-associated mode, respectively.
% For ports near the center of the chain, both amplitudes inherit the
% Majorana localization envelope,
% $|g_{\rm in}|,|g_{\rm out}|\sim e^{-N/\xi}$.
% Together with $\tau_{\rm e}\sim e^{N/\xi}$, this gives the
% protection--access law
% \begin{equation}
% R_{\rm loc}
% \sim e^{-2N/\xi}
% \sim \tau_{\rm e}^{-2}.
% \label{eq:protection_access}
% \end{equation}
% Figure~\ref{fig:access}(a) verifies this scaling for different
% topological parameters.
% Thus, the same localization that produces exponentially long edge
% memory also suppresses its local write--read access.

Long retention, however, does not necessarily imply strong local access.
For the central local input and a readout over a fixed finite set of bulk sites, we write the edge contribution to the impulse response as $|h_d^{(\mathrm e)}| =\mathcal{R}_{\rm loc}|\lambda_{\rm e}|^d$, where $\mathcal{R}_{\rm loc}$ measures the edge-response amplitude accessible through the local interface.
The relevant slow coordinate is the nonlocal Majorana parity $\ii\gamma_L\gamma_R$.
A local Majorana bilinear in the bulk couples to this coordinate through the product of the left and right Majorana tails, $|\ell_j r_j|\sim e^{-N/\xi}$.
The local write and read amplitudes therefore each scale as $e^{-N/\xi}$, giving the protection--access law
\begin{equation}
\mathcal{R}_{\rm loc}
\sim e^{-2N/\xi}
\sim c\,\tau_{\rm e}^{-2},
\label{eq:protection_access}
\end{equation}
where $c$ is a coefficient determined by the system size and model parameters
(see SM Sec.~S2~\cite{SM}).
Figure~\ref{fig:access}(a) verifies this scaling for different parameters in the topological phase.
% The same localization that gives the edge mode its exponentially long lifetime therefore makes it exponentially difficult to access through local bulk interfaces.

\begin{figure*}[t]
\centering
\includegraphics[width=0.98\textwidth]{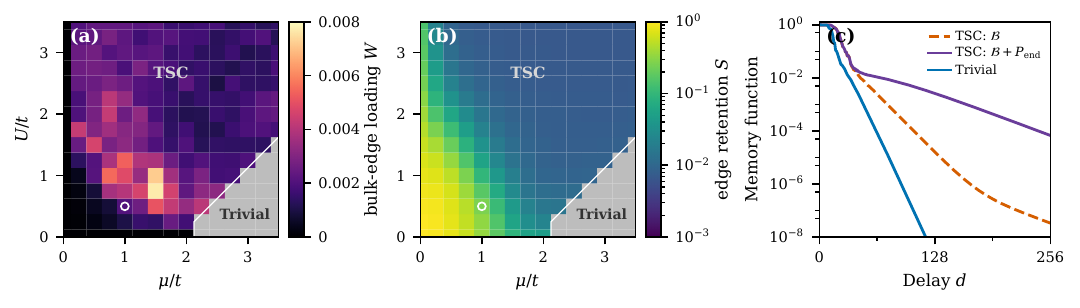}
\par\vspace{-3pt}

\includegraphics[width=0.98\textwidth]{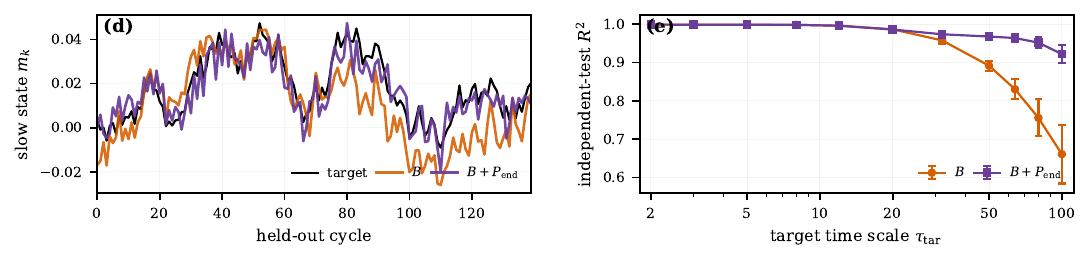}
\caption{\textbf{Interaction-induced memory consolidation and edge retention.}
(a) TSC and Trivial denote the topological superconducting and trivial phases, respectively.
The color scale shows the bulk--edge loading $W$ from the center-derived bulk into the edge mode.
Loading vanishes at $U=0$.
(b) Mean edge retention $S$ of the edge mode.
Interactions enable loading at the cost of edge retention.
The open circles in (a,b) mark the topological operating point used in (c--e).
(c) Memory function for local and edge-augmented readouts in the topological phase, with a trivial reference using $\mathcal B+P_{\rm end}$.
(d) Target and held-out predictions for a temporal integration task with
$\tau_{\rm tar}=100$, using the local features $\mathcal{B}$
and the augmented feature set $\mathcal{B}+P_{\rm end}$.
(e) Test $R^2$ as a function of the target memory timescale
$\tau_{\rm tar}$.
Parameters are $N=10$, $\Delta=t$, and $t\tau=1.5$,
with $\mu/t=1$ and $U/t=0.5$ for the topological phase in (c--e);
the trivial reference in (c) uses $\mu/t=3$.
% Loading and retention in (a,b) are evaluated over $D=160$.
}
\label{fig:interaction}
\end{figure*}

\textit{Programmable edge memory.—}
To overcome the local-access penalty, we keep the central input and introduce an ancillary fermion \(f_a\), reinitialized at each update, that couples directly to the nonlocal edge fermion \(f_{\rm e}=(\gamma_L+i\gamma_R)/2\).
The coupling is implemented by
$U_\theta=\exp[\theta(f_a^\dagger f_{\rm e} - f_{\rm e}^\dagger f_a)]$, where $\theta$ controls the exchange strength.
This additional input channel does not modify the underlying quantum dynamics of the Kitaev chain, but it changes the covariance-space dynamics. Eq.~\eqref{eq:linear_reservoir} therefore remains linear, while \(A\) and \(g\) become \(\theta\)-dependent. Since the ancillary channel is matched directly to the edge mode, the bulk sector is unchanged to leading order, and we focus on the resulting modification of the slow edge coordinate.
Writing the edge parity as $e_k=\langle i\gamma_L\gamma_R\rangle_k$, the resulting update is
\begin{equation}
e_{k+1}
=
\lambda_\theta e_k
+
\kappa_\theta u_k
+
O(e^{-N/(2\xi)}),
\label{eq:matched_edge_update}
\end{equation}
where $\lambda_\theta=\cos^2\theta$ and $\kappa_\theta=\sin^2\theta$ are the retention and write coefficients, respectively.
Unlike the exponentially suppressed local write amplitude, $\kappa_\theta$ remains finite as the system size increases.
The matched interface therefore turns the protected edge mode into a directly writable long-term memory register.

The write and retention coefficients satisfy $\kappa_\theta=1-\lambda_\theta$.
Defining the retention time as $\tau_\theta=-1/\ln\lambda_\theta$, we obtain $\tau_\theta\sim\kappa_\theta^{-1}$ for weak writing, $\kappa_\theta\ll1$.
Increasing $\theta$ writes a larger fraction of each new input into the edge memory, but also accelerates the decay of previously stored information.
The corresponding edge-memory function is $C_d^{(\mathrm e)}=(1-\lambda_\theta^2)\lambda_\theta^{2d}$, with $\sum_{d\geq0}C_d^{(\mathrm e)}=1$.
The edge mode provides one programmable memory direction whose weight across delays is controlled by $\theta$, adding a tunable slow memory channel to the faster, higher-dimensional bulk dynamics.
Figure~\ref{fig:access}(b) shows this write--retention tradeoff
(see SM Sec.~S3~\cite{SM}).

\textit{Interactions: bulk--edge loading and retention.—}
Turning on the interaction~\cite{gangadharaiah2011,katsura2015} enables autonomous transfer of information from the fast bulk into the long-lived edge. 
Once the interaction is introduced (\(U\neq0\) in Eq.~\eqref{eq:H}), the exact covariance-space description of the free reservoir no longer closes. 
This enables memory consolidation while simultaneously reducing edge retention, leading to a competition between memory loading and protected storage.

Decompose the Hamiltonian in Eq.~\eqref{eq:H} as $H=H_0+UV$, where
$H_0$ and $V$ are the free and interaction parts, respectively.
To isolate interaction-induced loading from the bulk into the edge, we
define a source $\Phi_q$ from the central input operator $P_q$.  It is
obtained by projecting $P_q$ onto the conserved free bulk-mode-parity
subspace and orthogonalizing the result against $P_{\rm end}$ with respect
to the Hilbert--Schmidt inner product
$(X|Y)=2^{-N}\mathrm{Tr}(X^\dagger Y)$.  Consequently,
$[H_0,\Phi_q]=0$ and $(P_{\rm end}|\Phi_q)=0$.
% Its explicit projector construction is given in the Supplemental Material.

To formulate the perturbative transfer, introduce the Liouvillian
superoperators $\mathcal L_0X=[H_0,X]$ and
$\mathcal L_{\rm int}X=[V,X]$.  Both $\Phi_q$ and $P_{\rm end}$ are
Majorana bilinears.  Free evolution generated by $\mathcal L_0$ preserves
the bilinear sector, whereas one action of $\mathcal L_{\rm int}$
maps a bilinear into the quartic sector.  Hilbert--Schmidt orthogonality of
these operator sectors eliminates the first-order contribution in $U$.
The leading interaction-induced transfer is therefore
\begin{equation}
(P_{\rm end}|
e^{-\ii(\mathcal L_0+U\mathcal L_{\rm int})s}
|\Phi_q)
=
U^2\Xi^{(2)}_{q\rightarrow{\rm end}}(s)+O(U^3).
\label{eq:second_order_loading}
\end{equation}
Here $\Xi^{(2)}_{q\rightarrow{\rm end}}(s)$ is the time-ordered second-order Dyson kernel with two insertions of $\mathcal L_{\rm int}$.
The first interaction vertex transfers the bulk bilinear into the
quartic sector, and the second generates the bilinear component detected by
$P_{\rm end}$. See SM Sec.~S4~\cite{SM} for detailed derivations.

To characterize the loading--retention competition at finite $U$ over $D$ cycles, we use the full closed-system one-cycle
propagator $\mathcal U_U=e^{-\ii(\mathcal L_0+U\mathcal L_{\rm int})\tau}$.
The mean squared endpoint autocorrelation defines the edge retention,
$S=D^{-1}\sum_{d=0}^{D-1}
|(P_{\rm end}|\mathcal U_U^d|P_{\rm end})|^2$, while the largest
bulk-to-edge transfer amplitude defines the loading,
$W=\max_{0\leq d<D}
|(P_{\rm end}|\mathcal U_U^{d+1}|\Phi_q)|$.
Within the finite-size topological superconducting (TSC) regime of
the interacting chain~\cite{miao2017,mahyaeh2020}, Fig.~\ref{fig:interaction}(a) shows that bulk--edge
loading develops a broad maximum at intermediate $U$ and $\mu$.
Figure~\ref{fig:interaction}(b) shows the complementary trend:
edge retention is largest deep in the TSC regime for weak interactions and
decreases as $U$ grows or the phase boundary is approached.
Together, these demonstrate the loading--retention tradeoff between
interaction-enabled memory consolidation and protected edge storage.
Figure~\ref{fig:interaction}(c) shows that the topological edge mode enhances the long-term memory, while leaving short-delay memory nearly unchanged.

We next show that interaction-loaded edge memory improves
long-timescale reservoir processing using a temporal
integration task.
For independent inputs $u_k\sim\mathcal{U}[-1,1]$, we define the target
\begin{equation}
m_k^{(\tau_{\rm tar})}
=
\alpha m_{k-1}^{(\tau_{\rm tar})}
+
(1-\alpha)u_k,
\qquad
\alpha=e^{-1/\tau_{\rm tar}},
\label{eq:temporal_integral}
\end{equation}
where $\tau_{\rm tar}$ sets the target memory timescale.
We compare a readout using the local bilinear feature set $\mathcal{B}$
with an augmented readout $\mathcal{B}+P_{\rm end}$.
As shown in Fig.~\ref{fig:interaction}(d,e), both readouts perform
similarly at short target timescales, whereas including $P_{\rm end}$
substantially improves performance at long timescales.
The interaction-loaded edge coordinate therefore extends the accessible
memory timescale beyond that provided by the local bulk features.
Numerical details are given in SM Sec.~S5~\cite{SM}.
We also demonstrate the system's performance on a nonlinear task and its robustness to disorder and local defects.
Detailed results are shown in SM Sec.~S6~\cite{SM}.

\textit{Discussion and conclusion.—}
% We have introduced topological quantum reservoir computing, in which the bulk--edge structure separates fast information processing from long-lived storage within a single quantum reservoir.
We have introduced topological quantum reservoir computing, in which the physical system supporting protected memory also performs computation through its bulk dynamics.
For the Kitaev chain, topological localization produces exponentially long edge retention while simultaneously suppressing local access, revealing a fundamental protection--access tradeoff.
A matched edge interface restores direct writing and makes the protected mode programmable, while turning on interactions enable autonomous bulk--edge memory consolidation at the cost of reduced retention.
These tradeoffs show that useful topological memory requires protection, access, and loading to be considered together. 
% In this architecture, topology sets the long memory timescale, the interface controls external access to the protected sector, and interactions control information transfer between the fast bulk and slow edge sectors.

In this architecture, topological localization supports long-lived edge memory, interfaces control access to that memory, and interactions mediate information transfer from the bulk to the edge.
% A further direction is to  engineer the input interfaces and interaction structure to achieve more flexible control of information access, transfer, and retention.
Further control of the interfaces and interaction structure could enable more flexible tuning of information transfer and retention.
% Although demonstrated here using the Kitaev chain, the underlying principle extends to other quantum systems with topological boundary modes, such as the Su–Schrieffer–Heeger chain~\cite{su1979ssh}, and even beyond quantum systems, since topological boundary modes also arise in classical physical systems~\cite{ma2019,ozawa2019}.
The same organizing principle may extend to other systems with topological boundary modes, such as the Su–Schrieffer–Heeger chain~\cite{su1979ssh}, and even to classical topological platforms~\cite{ma2019,ozawa2019}.
By combining bulk dynamics and protected edge memory in a single reservoir, TQRC establishes a new approach to quantum information processing and opens a route toward near-term quantum machine learning.

\textit{Acknowledgement.---}
This work was supported by Moonshot R\&D (Grant No. JPMJMS2021); a project JPNP14004, commissioned by the New Energy and Industrial Technology Development Organization (NEDO); JST NEXUS Japan (Grant No. JPMJNX26C5); and the National Natural Science Foundation of China (NSFC) (Grant No.12688201).

\bibliography{Topological_QRC}
\end{document}